% ``Direct Determination of DNA Twist-Stretch Coupling''
% Kamien Lubensky Nelson O'Hern
%
% print this using plain TeX

%SWITCHES
%\def\eplmode{T} % for submission to epl
\long\def\cut#1{}
%\long\def\optional#1{}
\def\pagin#1{}
\def\boringfonts{y}   % include for export
\def\figflag{y} \input epsf  % remove for submission

\input harvmac  %\input philmac

%%%%%%%%%%%%% philmac.tex %%%%%%%%%%%%%

\def\fonttest{y}
%macros useful in addition to mac.tex
% WARNING: what was once \fig is now \pfig !!
%           "    "   "   \listtoc "  \plisttoc
%           "    "   "   \nfig       \pnfig
%           "    "   "   \del        \pdel

% first a full nine-point font set
\ifx\boringfonts\fonttest
\else

\fi

\hyphenation{anom-aly anom-alies coun-ter-term coun-ter-terms
dif-feo-mor-phism dif-fer-en-tial super-dif-fer-en-tial dif-fer-en-tials
super-dif-fer-en-tials reparam-etrize param-etrize reparam-etriza-tion}

%\input usr1:[nelson.texutil]mac.tex

%
% Tagged sections: generates a symbol with current secno and also a
%               table of contents entry
%
% The page number in the toc may be wrong when a section contains no
% text. Try modifying \tnewsec by eliminating \let\the=0 .
%
\newwrite\tocfile\global\newcount\tocno\global\tocno=1
\ifx\bigans\answ \def\tocline#1{\hbox to 320pt{\hbox to 45pt{}#1}}
\else\def\tocline#1{\line{#1}}\fi
\def\toclead{\leaders\hbox to 1em{\hss.\hss}\hfill}
\def\tnewsec#1#2{\newsec{#2}\xdef #1{\the\secno}%
\ifnum\tocno=1\immediate\openout\tocfile=toc.tmp\fi\global\advance\tocno
by1%
{\let\the=0\edef\next{\write\tocfile{\medskip\tocline{\secsym\ #2\toclead\the%
\count0}\smallskip}}\next}% want to expand secsym now, count0 later
\nobreak}
\def\tnewsubsec#1#2{\subsec{#2}\xdef #1{\the\secno.\the\subsecno}%
\ifnum\tocno=1\immediate\openout\tocfile=toc.tmp\fi\global\advance\tocno
by1%
{\let\the=0\edef\next{\write\tocfile{\tocline{ \ \secsym\the\subsecno\
#2\toclead\the\count0}}}\next}%
\nobreak}
\def\tappendix#1#2#3{\xdef #1{#2.}\appendix{#2}{#3}
\ifnum\tocno=1\immediate\openout\tocfile=toc.tmp\fi\global\advance\tocno
by1
{\let\the=0\edef\next{\write\tocfile{\tocline{ \ #2.
#3\toclead\the\count0}}}\next}
}
%
% generate rudimentary table of contents
%

%
%
%
% manuscript control
%
% Phys Rev Letters: (those jerks)
% doublespace, put footnotes among references, doublespace refs, set flag

\gdef\prlmode{F}
%
% Science: (those jerks)

%
\long\def\optional#1{}
\def\cmp#1{#1}         %or remove for stuffy journals
%
%   Matters of taste
%
\let\narrowequiv=\equiv
\def\equiv{\;\narrowequiv\;}

\def\tilde{\widetilde}
     %or else other way round
\fontdimen16\tensy=2.7pt\fontdimen17\tensy=2.7pt %an experiment
 %usual dup not needed with the above
%\mathsurround=1pt %screws up \dsl!

% for the drafts this is useful: (remember not to do \listrefs)
%\def\ref#1#2{\edef\tmp{$[$\string#1$]$} \tmp\edef#1{\tmp}}

% for reduction this is useful:
% \ifx\answ\bigans ...<some equation> \else ... <same with linebreaks> \fi

%
%Greek abbreviations

\def\al{{\alpha}}

%
%Curly letters
%

%
%
%   Miscellaneous
%
% Usage: \boxit{3.5}{some text}
\def\boxit#1#2{
        $$\vcenter{\vbox{\hrule\hbox{\vrule\kern3pt\vbox{\kern3pt
	\hbox to #1truein{\hsize=#1truein\vbox{#2}}\kern3pt}\kern3pt\vrule}
        \hrule}}$$
}

%\def\bivector#1{{\buildrel \leftrightarrow\over #1}}

%\def\lsim{\protect\raisebox{-0.75ex}[-1.5ex]{$\;\stackrel{<}{\sim}\;$}} latex

%\def\gsim{\protect \raisebox{-0.75ex}[-1.5ex]{$\;\stackrel{>}{\sim}\;$}}%latex

         % |#1>
         % <#1|
 %matrix element <#1|#2|#3>
  %bold nabla
%

 %       little fraction
 %     tensor up-down

%\def\THETA#1#2#3{\vartheta \hbox{${#1\atopwithdelims[]#2}$}
%               \bigl(#3|\tau\bigr)}
%\def\THETB#1#2#3#4{\vartheta \hbox{${#1\atopwithdelims[]#2}$}
%               \bigl(#3|#4\bigr)}

             % no dot i

% split exact sequence
\def\splitexact#1#2{\mathrel{\mathop{\null{
\lower4pt\hbox{$\rightarrow$}\atop\raise4pt\hbox{$\leftarrow$}}}\limits
^{#1}_{#2}}}

%semi-direct product |><
%
%   Various delbars
%

            %delbar dagger
              %delbar dagger delbar
  %delbar n dagger delbar n
            %del-bar-sub n
           %delbar
 %partial derivative
 %partial derivative with room for tilde etc
     %d'alembertian
%
%   Other things with bars
%
          % Beltrami
 % conjugate Beltrami
  % z-bar
  % q-bar
  % tau-bar
  % u-bar
  % a-bar
      % mu-bar
%
%   Various romans for math
%

  % Im, Re
\def\ex#1{{\rm e}^{#1}}                 % exponential
\def\dd{\mskip 1.3mu{\rm d}\mskip .7mu} % exterior derivative
                   % dz dzbar
                      % trace

        % DET bundle
        % DET bundle

                % det
                % sdet

%
%   Whole words
%

\def\IM{isomorphism}

%
%   Fonts
%

%\def\cmfontflag{cm}
%\ifx\fontflag\cmfontflag\font\CAPS=cmcsc10 scaled 1200
% \ifx\answ\bigans\font\speci=bbb12 scaled 833\else\font\speci=bbb10\fi
%\else
% \font\CAPS=amcsc10 scaled 1200
%\fi
\ifx\boringfonts\fonttest
\font\blackboard=cmssbx10 \font\blackboards=cmssbx10 at 7pt  % wimpy
\font\blackboardss=cmssbx10 at 5pt
\else
\font\blackboard=msym10 \font\blackboards=msym7   % cool
\font\blackboardss=msym5
\fi
\newfam\black
\textfont\black=\blackboard
\scriptfont\black=\blackboards
\scriptscriptfont\black=\blackboardss

%\def\spec#1{\hbox{\speci #1}}
		      %historical

%
\ifx\boringfonts\fonttest
\font\gothic=cmssbx10 \font\gothics=cmssbx10 at 7pt  % wimpy substitute
\font\gothicss=cmssbx10 at 5pt
\else
\font\gothic=eufm10 \font\gothics=eufm7
\font\gothicss=eufm5
\fi
\newfam\gothi
\textfont\gothi=\gothic
\scriptfont\gothi=\gothics
\scriptscriptfont\gothi=\gothicss

{\catcode`\@=11\gdef\oldcal{\fam\tw@}}
\newfam\curly
\ifx\boringfonts\fonttest\else
\font\curlyfont=eusm10 \font\curlyfonts=eusm7
\font\curlyfontss=eusm5
\textfont\curly=\curlyfont
\scriptfont\curly=\curlyfonts
\scriptscriptfont\curly=\curlyfontss

\fi
%

%\font\bbol=cmbx10 scaled \magstep1    %chapter titles
\ifx\boringfonts\fonttest\def\df{\bf}\else\font\df=cmssbx10\fi

%macros to get automatically-numbered figures
\global\newcount\pnfigno \global\pnfigno=1
\newwrite\ffile
\def\pfig#1#2{Fig.~\the\pnfigno\pnfig#1{#2}}
\def\pnfig#1#2{\xdef#1{Fig. \the\pnfigno}%
\ifnum\pnfigno=1\immediate\openout\ffile=figs.tmp\fi%
\immediate\write\ffile{\noexpand\item{\noexpand#1\ }#2}%
\global\advance\pnfigno by1}

% This one embeds figs in the text (don't mix \tfig with \pfig!)
% Unlike \pfig, this one comes in two parts: \tfig allocates the number and
% \ifig inserts the empty box
% e.g. blah, blah (\tfig\foo) blah, end of paragraph.
%
% \ifig\foo{caption.}{2.5}
%
% New paragraph; a box 2.5truein has been left and captioned.
\def\tfig#1{Fig.~\the\pnfigno\xdef#1{Fig.~\the\pnfigno}\global\advance\pnfigno
by1}

% here come Distler's versions
% Interface to epsf.tex
% the call (after assigning a figure number with \tfig) is
%
%\ifigure\figlabel{caption}{figfile}{vsize}
%
%If \figflag is undefined, it leaves a vbox with caption and .2truein
%of space (like \ifig).  If you have \def\figflag{y}, it inserts the figure
%scaled to fit. The logic here is that you can turn off all figure
%generation by commenting out one line at the beginning of the file
%(the line that \def's \figflag). The pagination, etc. is completely
%the same whether the figure is there or not.
%
%There is also a macro called \epsfsize, which checks that the figure is
%narrow enough to fit in the current \hsize. If not, it is again scaled to fit.
%
\def\figI{y}
\def\ifigure#1#2#3#4{
\midinsert
\ifx\figflag\figI
 \ifx\htflag\figI
 \vbox{
  \href{file:#3}% assumes figure has been loaded locally (no absolute
	       % url capability in xhdvi); useful for a better look
{Click here for enlarged figure.}}
 \fi
 \vbox to #4truein{
 \vfil\centerline{\epsfysize=#4truein\epsfbox{#3}}}
\else
\vbox to .2truein{}
\fi
\narrower\narrower\noindent{\bf #1:} #2
\endinsert
}

% new stuff for bozo

%\def\hal{{1\over2}} %obsolete

%\def\sump{\CO({\textstyle{\sum_{i=1}^q}}Q_i-{\textstyle{\sum_{j=1}^q}}P_i)}

%%%%%%%%%%%%%% goodies for super riemann surfaces %%%%%%%%%%%%%%

 %beltrami diffl

   % big D

                     % bold z
                     % bold u
  % bold z-bar
  % bold u-bar

%
% A-hats
   %curly A-hat
   %curly A-hat, holo
   %curly A-hat, antiholo
%
% Omega-hats

%
% O-hats

%%%%%%%%%%%%%%%%%%%%%%%%%%%%%%%%%%%%%%%%

% Poor man's Blackboard Bold characters often used :
\def\inbar{\,\vrule height1.5ex width.4pt depth0pt}
\def\IB{\relax{\rm I\kern-.18em B}}
\def\IC{\relax\hbox{$\inbar\kern-.3em{\rm C}$}}
\def\ID{\relax{\rm I\kern-.18em D}}
\def\IE{\relax{\rm I\kern-.18em E}}
\def\IF{\relax{\rm I\kern-.18em F}}
\def\IG{\relax\hbox{$\inbar\kern-.3em{\rm G}$}}
\def\IH{\relax{\rm I\kern-.18em H}}
\def\II{\relax{\rm I\kern-.18em I}}
\def\IK{\relax{\rm I\kern-.18em K}}
\def\IL{\relax{\rm I\kern-.18em L}}
\def\IM{\relax{\rm I\kern-.18em M}}
\def\IN{\relax{\rm I\kern-.18em N}}
\def\IO{\relax\hbox{$\inbar\kern-.3em{\rm O}$}}
\def\IP{\relax{\rm I\kern-.18em P}}
\def\IQ{\relax\hbox{$\inbar\kern-.3em{\rm Q}$}}
\def\IR{\relax{\rm I\kern-.18em R}}
\font\cmss=cmss10 \font\cmsss=cmss10 at 10truept%!!! should be 7pt
\def\IZ{\relax\ifmmode\mathchoice
{\hbox{\cmss Z\kern-.4em Z}}{\hbox{\cmss Z\kern-.4em Z}}
{\lower.9pt\hbox{\cmsss Z\kern-.36em Z}}
{\lower1.2pt\hbox{\cmsss Z\kern-.36em Z}}\else{\cmss Z\kern-.4em Z}\fi}
\def\IGa{\relax\hbox{${\rm I}\kern-.18em\Gamma$}}
\def\IPi{\relax\hbox{${\rm I}\kern-.18em\Pi$}}
\def\ITh{\relax\hbox{$\inbar\kern-.3em\Theta$}}
\def\IOm{\relax\hbox{$\inbar\kern-3.00pt\Omega$}}

\def\exv#1{{\langle #1\rangle}}
\def\ee#1{\cdot 10^{#1}}

\def\const{{\rm const.}}

\def\tot{{\rm tot}}

% LREFS
\lref\rRudnick{B. Fain,
J. Rudnick, and S. \"{O}stlund, \cmp{``Conformations of linear DNA,''}
preprint (1996). }
\lref\rKrenk{For a detailed discussion see
S. Krenk, \cmp{``A linear theory of pretwisted elastic
beams,''} J. Appl. Mech. {\bf50} (1983) 137\pagin{--142}.}
\lref\rBenh{See for example C. Benham, \cmp{``Geometry and mechanics
of DNA superhelicity,''} Biopolymers {\bf22} (1983)
2477\pagin{--2495}.}
\def\rsuper{\rBenh}\def\rlink{\rBenh}
\lref\olddna{M.H.F. Wilkins, R.G. Gosling, and W.E. Seeds, Nature
{\bf167} (1951) 759.}
\lref\rRecord{See M. Record, S. Mazur, P. Melancon, J. Roe, S. Shaner, and
L. Unger,
\cmp{``Double helical DNA: conformations, physical properties, and
interactions with ligands,''} Annu. Rev. Biochem. {\bf50} (1981)
997\pagin{--1024}, and references therein.\optional{[old measurements
of A, B] }}
\lref\rfirst{S.B.~Smith, L.~Finzi and C.~Bustamante,
      \cmp{``Direct mechanical measurements of the elasticity of
single DNA molecules       by using magnetic beads,''}
    Science {\bf 258} (1992) 1122\pagin{--6};
C.~Bustamante, J.F.~Marko, E.D.~Siggia and S.~Smith,
     \cmp{``Entropic elasticity of lambda-phage DNA,''}
  Science {\bf 265} (1994) 1599\pagin{--600}.
\optional{[early tweezer measurements of A]}}
\lref\rMarkonew{J. Marko, ``Stretching must twist DNA,'' preprint
1996.}
\lref\rLandau{L. Landau and E. Lifshitz, {\sl Theory of elasticity}
3$^{rd}$ ed.  (Pergamon, 1986). Strictly speaking the elastic constants
in \etsbm\  differ slightly from the measured values
we use, but
for our estimate we will neglect this correction.}
\lref\rCalla{C. Calladine and H. Drew, {\sl Understanding DNA} (Academic,
1992).}
\lref\rMSb{J.F.~Marko
and E.D.~Siggia, ``Stretching DNA,'' Macromolecules {\bf 28} (1995)
8759\pagin{--8770}.}
\lref\rMSa{J.F.~Marko
and E.D.~Siggia, ``Bending and twisting elasticity of DNA,''
Macromolecules, {\bf 27} (1994) 981\pagin{--988}.}
\def\rsecond{\rSmCuBu\rBlock}
\lref\rSmCuBu{S. Smith, Y. Cui, and
C. Bustamante, \cmp{``Overstretching B-DNA: the elastic response of
individual double-stranded and single-stranded DNA molecules,''}
Science {\bf271} (1996) 795\pagin{--799}.}
\lref\rBlock{M.D. Wang, H. Yin, R. Landick, J. Gelles,
and S.M. Block,
\cmp{``Stretching DNA with optical tweezers,''}
Biophys. J., in press (1997).}
\lref\rCluz{P. Cluzel, A. Lebrun, C. Heller, R. Lavery,
J.-L. Viovy, D. Chatenay, and F. Caron, \cmp{``DNA: an extensible
molecule,''} Science {\bf271} (1996) 792\pagin{--794}.}
\lref\rStrick{T. Strick, J. Allemand, D. Bensimon, A. Bensimon, and V.
Croquette, \cmp{``The elasticity of a single supercoiled DNA
molecule,''}
Science {\bf271} (1996) 1835\pagin{--1837}.}

%%%%%%%%%%%%%%%%%%%%%%%%%%%%%%%%%%%%%%%%%%%%%%%%%%%%%%%%%%%%%%%%%%%%%%
% HACKS

\def\cbend{A}%{{\hbox{\df A}}}% bend
\def\ctwist{C}%{{\hbox{\df C}}}% twist
\def\cstretch{B}%{{\hbox{\df B}}}% stretch
\def\cts{D}%{{\hbox{\df D}}}%twist-stretch
\def\ctb{G}%{{\hbox{\df G}}}%twist-bend
\def\cbs{K}%{{\hbox{\df K}}}%bend-stretch
\def\exv#1{{\langle #1\rangle}}
\let\epsilon=\varepsilon
\let\phi=\varphi
\def\ddd#1#2{{{\rm d}#1\over{\rm d}#2}}
 
\def\const{{\rm const.}}

\def\tot{_{\rm tot}}

\def\lk{{\rm Lk}}

\def\wo{{\omega_0}}
\def\do{{d_0}}
\def\EE{\hat E_1}\def\FF{\hat E_2}\def\TT{\hat E_3}
\def\GG#1{\hat E_{#1}}
\def\EEZ{\hat E_{10}}\def\FFZ{\hat E_{20}}\def\TTZ{\hat E_{30}}
\def\ee{\hat e_1}\def\ff{\hat e_2}
\def\OO#1{{\Omega_{#1}}}
\hfuzz=3truept

%\let\lref=\ref
%%%%%%%%%%%%%%%%%%%%%%%%%%%%%%%%%%%%%%%%%%%%%%%%%%%%%%%%%%%%%%%%%%%%%%

% BEGINNING

\def\testp{T}

\ifx\eplmode\testp
\def\newsec#1{\global\advance\secno by1\message{(\the\secno. #1)}
{\sl \the\secno. #1:\ \ }}
\fi

\Title{UPR--725T}{\vbox{\centerline{Direct Determination of DNA }
\vskip2pt\centerline{Twist-Stretch Coupling}
}}

\centerline{Randall D. Kamien, Tom C. Lubensky, Philip Nelson, and
Corey S. O'Hern}\smallskip
\centerline{Department of Physics and
Astronomy,}\centerline{University of Pennsylvania,
Philadelphia, PA 19104 USA}
\bigskip

The symmetries of the DNA double helix require a new term in its
linear response to stress: the coupling between twist and stretch.
Recent experiments with torsionally-constrained single molecules give
the first direct measurement of this important material parameter. We
extract its value from a recent experiment of Strick {\it et al.} [Science
{\bf271} (1996) 1835] and find rough agreement with an independent
experimental estimate recently given by Marko. We also present a very
simple microscopic theory predicting a value comparable to the one
observed.

\noindent {\sl PACS:
87.15.-v, %  Molecular biophysics
87.10.+e, %  General, theoretical, and mathematical biophysics
87.15.By.%  Structure, bonding, conformation, configuration, and isomerism of
% biomolecules
}

\Date{{\sl11/96}}\noblackbox
%\draft\nolabels

\newsec{Introduction}
The idea of studying the response of DNA to mechanical stress
%treating DNA by the methods of classical elasticity theory
is as old as the discovery of the double helix structure itself
\olddna. While many elements of DNA function require detailed
understanding of specific chemical bonds (for example the binding of
small ligands), still others are quite nonspecific and reflect
overall mechanical properties. Moreover, since
the helix repeat distance of $\ell_0\approx3.4\,$nm involves dozens of
atoms, it is reasonable to hope that this length-scale regime would be
long enough so that the cooperative response of many atoms would
justify the use of a continuum, classical theory, yet short enough
that the spatial structure of DNA matters. In this Letter we will
argue that this expectation is indeed fulfilled.

Since moreover various important biological processes involve length
scales comparable to $\ell_0$ (notably the winding of DNA onto
histones), the details of this elasticity theory should prove
important. Yet until recently little was known about the relevant
elastic constants. Extensive experimental work yielded fair agreement
on the values of the bend and twist persistence lengths, though the
former was plagued with uncertainties due to the polyelectrolyte
character of DNA \rRecord. A simple model of DNA as a circular elastic
rod % with these moduli
gives a reasonable account of many features of
its long-scale behavior, for example supercoiling
\rsuper\rRudnick. Some
authors have sought to justify this model by invoking a shell of
structured water around the DNA \rBenh.

Recently, techniques of micromanipulation via optical tweezers and
magnetic beads have yielded improved values for the bend stiffness from the
phenomenon of thermally-induced entropic elasticity \rfirst\rMSb, as
well as the direct measurement of a third elastic constant, the
stretch modulus, by exploring the force range 10--50pN
\rsecond. Significantly, the relation between bending stiffness,
stretch modulus, and the
diameter of DNA turned out to be roughly as predicted from the
classical theory of beam elasticity \rsecond\rLandau, supporting the
expectations mentioned above.

Still missing, however, has been any direct measurement of the elastic
constants reflecting the {\it chiral} ({\it i.e.} helical) character
of DNA. One such constant, a twist-bend coupling, was investigated by
Marko and Siggia \rMSa, but no direct experimental measurement has yet
been devised. In this Letter we introduce a new chiral coupling, the
twist-stretch energy. Electrostatic effects do not complicate the
analysis of this coupling.  We will explain why our term is needed,
extract its value from the experiment of Strick {\it et al.} \rStrick,
and compare it to a the prediction of a
simple microscopic model to see that its magnitude is
in line with the expectations of classical elasticity theory. J. Marko
has independently introduced the same coupling and estimated its value
from different experiments\rMarkonew; our values are in rough
agreement.

\newsec{Experiment}
DNA differs from simpler polymers in that it can resist twisting, but
it is not easy to measure this effect directly due to the difficulty of
applying external torques to a single molecule.
Early  investigations of DNA twist were either limited to passive,
fluorescence-depolarization measurements  \rRecord, or else to
studying global shape changes in circular DNA of varying linkage
\rlink.  The first single-molecule stretching experiments constrained
only the locations of the two ends of the DNA strand.
The unique feature of the experiment of Strick {\it et al.}  was the
added ability
to constrain the {\it orientation} of each end of the molecule.

We will study Fig.~3a of ref.~\rStrick. In this experiment, a constant
force of 8pN was applied to the molecule and the end-to-end length
$z\tot$ monitored as the terminal end was rotated through $\Delta\lk$
turns from its relaxed state (which has $\lk_0$ turns). In this way the
helix could be over- or undertwisted by as much as $\pm10$\%. Over
this range of imposed linkage $z\tot$ was found to be a linear function
of $\sigma$:
\eqn\eexpt{\epsilon=\const-0.15\sigma\, {\rm\  where\ \ }
\sigma\equiv\Delta\lk/\lk_0\ {\rm\  and\ \ }
\epsilon\equiv(z\tot/z_{{\rm tot,}0})-1\ .}
Thus $\sigma$ is the fractional excess link and $\epsilon$ is the
extension relative to the relaxed state.
Eqn.~\eexpt\ is the experimentally
observed twist-stretch coupling.

The existence of a linear term in \eexpt\ is direct evidence of the
chiral character of the molecule, and its sign is as expected on
geometrical grounds: untwisting the molecule tends to lengthen it.
Still geometry alone cannot explain this result. Consider the outer
sugar-phosphate backbones of the DNA. Suppose that the twist-stretch
phenomenon were due to the straightening of these helical backbones
while they
maintained constant length, 0.6~nm per  phosphate, and constant
distance 0.9~nm from the center of the molecule. Then since  each
basepair step is $h=0.34\,$nm high, the circumferential length per step
is $\ell_c=\sqrt{.6^2-.34^2}\,$nm. The
corresponding twist angle per step is given by \rCalla\
$\theta= (\ell_c/2)/.9{\rm nm}=32^\circ$, roughly as observed.
Supposing now an extension by $\Delta h/h=\epsilon$, we find an
untwisting by $\sigma=\delta\theta/\theta=\const-\epsilon/2.0$, quite
different from what is observed, eqn.~\eexpt. We must seek an
explanation of the experimental result not in terms of a geometrical
ball-and-stick model but in the context of an elastic response theory.

\newsec{Simple Model}
We will begin by neglecting bend fluctuations (see below). A straight
rod %of length $z\tot$
under tension and torque will stretch and twist. We
can describe it by the reduced elastic free energy
\eqn\ets{f_1(\sigma,\epsilon)\equiv{F_1(\sigma,\epsilon)
\over k_BTz_{\rm tot,0}}
={\wo^2\over2}\left[
\bar\ctwist\sigma^2+\bar\cstretch\epsilon^2
+2\bar\cts\epsilon\sigma\right]
 -\tau\epsilon\ .}
The twist persistence length is
$\bar\ctwist\approx 75\,$nm \rRecord, while the helix parameter
$\wo=2\pi/\ell_0=1.85/$nm.
We will take\foot{$\bar\cstretch$
reflects the intrinsic stretchiness of DNA, since electrostatic
self-repulsion simply shifts the equilibrium length without affecting
the spring constant. Indeed experiments show little or no dependence
of $\bar\cstretch$ on salt, unlike the situation with the effective bend
persistence length
\rBlock. We also expect $\bar\ctwist$ to reflect intrinsic elasticity, since
twisting does not affect the average charge distribution. }
$\bar\cstretch\approx1100\,{\rm pN}/\wo^2k_BT\approx 78\,$nm
\rBlock. In the
experiment under study the reduced force is $\tau=8\,$pN/$k_BT\approx
1.95/$nm. For a circular beam made of isotropic material the cross-term
$\bar\cts$ is absent \rLandau, since twisting is odd under spatial
inversion while stretching  is even. For a helical beam, however, we
must expect to find this term.

We now minimize $f_1$ with respect to $\epsilon,\ \sigma$ at fixed
tension with an imposed constraint on the overtwist
$\sigma$.
%Introducing a Lagrange multiplier $\Lambda$ (a torque), we let $f_1'=f_1-
%\Lambda\sigma$ and
Minimizing at fixed $\sigma$ and $\tau$ gives
$\Lambda=\wo^2(\bar\cts\epsilon +\bar\ctwist\sigma )$ and hence
\eqn\esolna{\epsilon=\epsilon_{\sigma=0}
-{(\bar\cts/\bar\cstretch)}\sigma\ .}
Comparing to \eexpt, we obtain the desired result: $\bar\cts=12\,$nm. To
compare this to Marko's analysis, we note that his dimensionless $g$
equals our
$\bar\cts\wo$, so that we get $g=22$. The rough agreement with Marko's
result $g=35$ \rMarkonew\ indicates that the data show a real
material parameter of DNA and not some artifact. We do not expect
exact agreement, since Marko %applied linear twist-stretch coupling to
studied
the nonlinear overstretching transition of \rCluz\rSmCuBu; our value
came from the linear regime of small strains.

\newsec{Bend Fluctuations}
To arrive at \ets\ we listed the variables which were constrained,
coupled to external forces, and/or observed in the experiment, namely
$\epsilon$ and $\sigma$, then wrote the most general quadratic
function allowed by symmetry. Thus \ets\ is a phenomenological model;
its coefficients $\bar\ctwist,\ \bar\cstretch,\ \bar\cts$ reflect both
intrinsic elasticity and the effects of thermal fluctuations. Indeed
it is well known that thermal bend fluctuations reduce the effective
stretch modulus at modest tension via the ``entropic elasticity''
effect \rfirst\rMSb. Somewhat inconsistently we arrived at our value
of $\bar\cts$ by using the intrinsic stretch modulus in \esolna. In
this section we will justify the procedure by introducing a more
elaborate model with bend fluctuations, rederiving the analog of
\esolna, and again comparing to \eexpt.

We begin by defining local variables (\tfig\fone a) (see
\rMSa\rCalla). DNA is a stack of base pairs. We neglect sequence
effects and so regard all base pairs as copies of one standard
slice. The standard slice contains a reference point with the property
that the locus of these is the helix axis, a straight line of length
$L$ in the relaxed state. Through this reference point we next draw a
fixed vector; a convenient choice is the ``dyad'' pointing into the
minor groove and perpendicular to the helix axis.

\ifigure\fone{Schematic diagrams defining  variables used in the
text. The offset from the helix axis has been exaggerated for clarity.
a)~Notation used in the study of bend fluctuations.  We
describe the DNA by the helix axis (dotted curve) and
the axis $\hat E_1$, which is a fixed vector in each base pair.
b)~Notation used in the microscopic model. The helix axis (dotted
line) is now supposed straight. We describe the
DNA by the dashed curve and the axis $\hat E_1$ as before.
}{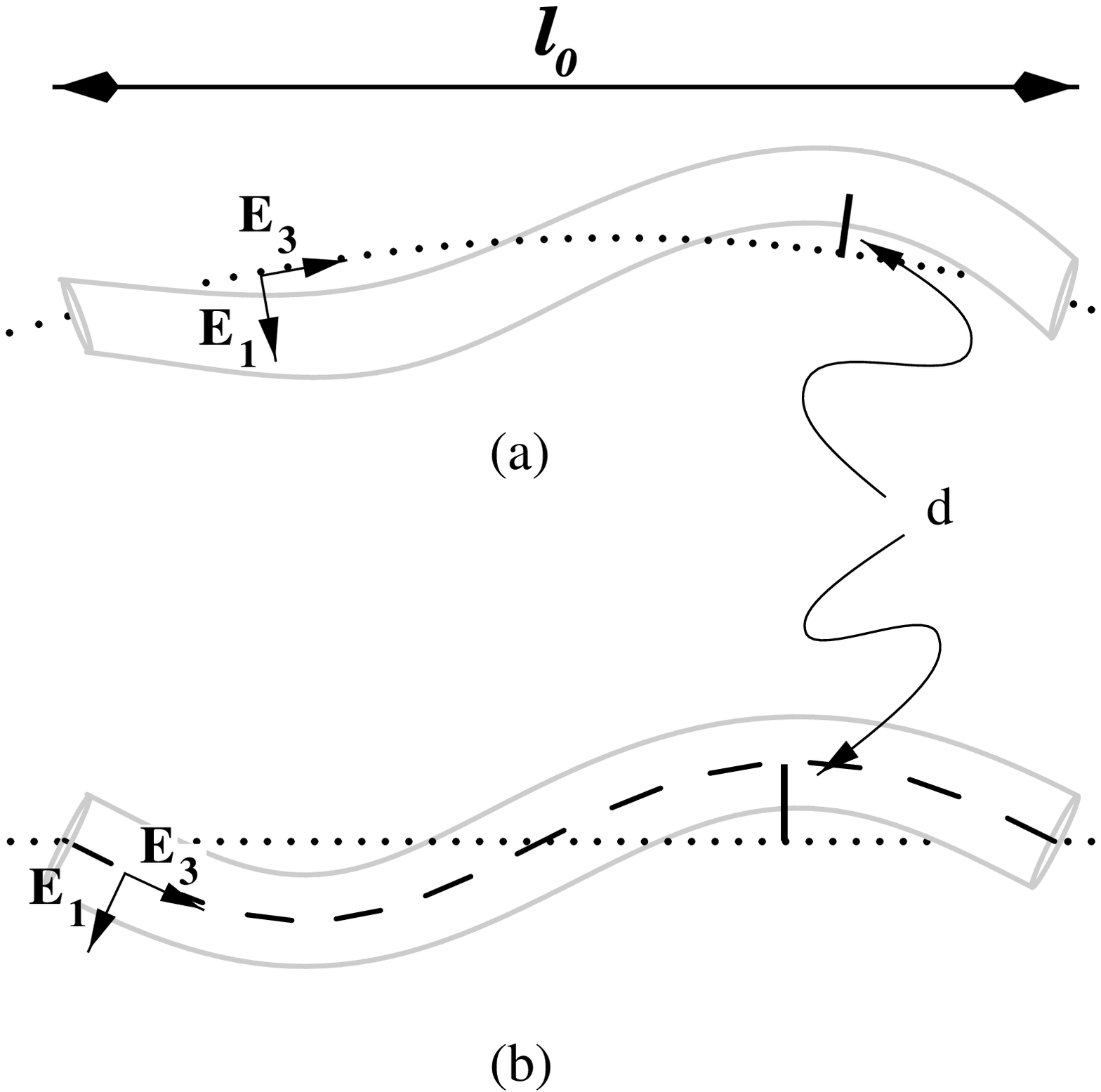}{3.5}

To describe stressed states, we simply specify the locus of reference
points as a parameterized curve in space (dotted line in \fone a) and
the dyad as a field of vectors $\EE$ normal to this curve. We let
$\TT$ be the unit tangent to the axis and complete to an orthonormal
triad by defining $\FF=\TT\times\EE$. Next we introduce a parameter
$s$ to label each slice; $s$ corresponds to arc length along the
original, unstressed helix axis and so always runs from 0 to $L$. The
actual arc length along the distorted axis will not be d$s$ but rather
$(1+\alpha(s))\dd s$;  $\alpha$ is thus the axial strain.

Thus our local variables are $\GG i(s)$ and $\alpha(s)$. Our program
consists of four steps: {\sl i)}~Find the strains in terms of the local
variables. {\sl ii)}~Write the
general linear elasticity theory of these strains with a force
coupling to the extension $\epsilon$ and a torque coupling to the
twist $\sigma$. {\sl iii)}~Compute and minimize the free energy to find the
end-to-end length $\exv{z\tot}$ in terms of the constrained $\sigma$
and the applied force $\tau$.  (It will prove convenient in this step
to convert to new variables
$t_1(s),\ t_2(s),\ \phi(s)$, and $\alpha(s)$.)
{\sl iv)}~We will then be able to relate the
experimental result to intrinsic elastic constants. We will see that
our naive calculation of the previous section is justified. Details of
this calculation will appear elsewhere.

{\sl Step i:}\/ In the relaxed state each slice bears a constant
relation to its predecessor. Thus while $\EEZ,\ \FFZ$, and $ \TTZ$
all vary in space, the derivatives with respect to $s$ are of
the form
\eqn\edfo{\dd{\hat E}_{i0}/\dd s=-\epsilon^{ijk}\Omega_{j0}\hat E_{k0}\ ,}
where $\Omega_{j0}$ are the constants $(0,0,\wo)$. For the deformed
state, \edfo\ defines the functions $\Omega_i(s)$. Our strain variables
are then
$\Omega_1(s),\ \Omega_2(s),\ \Omega_3(s)-\wo$, and $\alpha(s)$.

{\sl Step ii:}\/ Our strain variables have the desirable property that
in terms of them the elastic constants are independent of
$s$. Moreover the end-for-end symmetry of DNA implies that the elastic
matrix is unchanged upon changing the sign of $\Omega_1$ \rMSa. Thus we
generalize the model of \rMSa\ to
\eqn\ebt{\eqalign{f_2={1\over2L}\int_0^L\dd s\Bigl[&
\cbend'\OO1^2 + \cbend\OO2^2 + \ctwist(\OO3-\wo)^2 +
\cstretch\wo^2\alpha^2 + 2\cts\wo(\OO3-\wo)\al \cr&
+ 2\ctb(\OO3-\wo)\OO2 + 2\cbs\wo\OO2\alpha
\Bigr]\ .\cr}}
Here $\ctb$ is the twist-bend coupling of \rMSa, while $\cbs$ is an
allowed coupling between stretch and bend.
\cut{. The coupling $\cbs$
reflects the possibility that under extension the helix axis can move
away from the chosen reference point, so that the latter no longer
follows a straight line.}

We will apply a perturbative treatment to \ebt. Such an approximation
is valid since in the experiment we are analyzing the applied force is
large enough to keep the  end-to-end
distance over 90\% the full relaxed contour length, but not large
enough to create large intrinsic stretch $\alpha$. In addition,
the applied overtwist
$\sigma$ is at most $\pm10$\% \rStrick; indeed the slope reported in
\eexpt\ can be found from an even smaller range of $\sigma$ than
this.

{\sl Step iii:}~It proves  useful to refer the
frame $\{\EE,\ \FF,\ \TT\}$ to a  reference frame $\{\ee,\ \ff,\
\hat z\}$ rotating at $\wo$:
\ifx\eplmode\testp
$\ee=\hat x\cos(\wo s)+\hat y\sin(\wo s)$,
$\ff=-\hat x\sin(\wo s)+\hat y\cos(\wo s)$.
\else
\eqn\eframa{\ee=\hat x\cos(\wo s)+\hat y\sin(\wo s)\quad
\ff=-\hat x\sin(\wo s)+\hat y\cos(\wo s)\ .}
\fi
We will then write the deformed frame in terms of three small quantities: two
deviations of the tangent vector $t_{1,2}$ and an angle $\phi$. To
first order in these we have
\eqn\eframb{ \EE=\ee+\phi\ff-t_1\hat z\ ,
\quad \FF=-\phi\ee+\ff-t_2\hat z\ ,
\quad \TT=t_1\ee+t_2\ff+\hat z\quad\ .}
Thus $\EE$ is the unit vector perpendicular to the tangent whose
projection to the  $xy$ plane makes angle $\phi$ with the reference
frame. The advantage of these coordinates is now that the
excess link is simply the difference $\phi(L)-\phi(0)$. The relation
between Link, Twist, and
Writhe will also emerge automatically instead of having to be enforced
by hand.

Exercising the definitions we find that in terms of $t_1,\ t_2,\ \phi$
the strains are $\Omega_1=-\dot t_2-\wo t_1$, $\Omega_2=\dot t_1-\wo
t_2$, $\Omega_3=\wo+\dot \phi$. Substituting into \ebt, adding
external tension $\tau$ and torque $\Lambda$, and
rearranging gives the free energy
\eqn\ebtt{\eqalign{f_2={1\over2L}\int\dd s\,\biggl[&
\cbend'(\dot t_2+\wo t_1)^2
+ \cbend\Bigl(\dot t_1-\wo t_2+{\ctb\over\cbend}\dot\phi
          + {\cbs\wo\over\cbend}\alpha\Bigr)^2
\cr&
+\Bigl(\ctwist-{\ctb^2\over\cbend}\Bigl){\dot\phi}^2
+\wo^2\Bigl(\cstretch-{\cbs^2\over A}\Bigr)\alpha^2
+2\wo\Bigl(\cts-{\ctb\cbs\over\cbend}\Bigr)\dot\phi\alpha
\biggr]\cr
-{\tau\over L}\int\dd s\,(&1+\alpha)\bigl(1-\half {t_1}^2-\half{t_2}^2\bigr)
-{\Lambda\over L}\int\dd s\,\dot \phi
\ .\cr}}

{\sl Step iv:}~Let us first focus only on the linear couplings to the
applied forces
$\tau,\ \Lambda$. Then
eqn.~\ebtt\ shows that the Lagrange multiplier $\Lambda$ couples only
to the constant part of $\dot\phi$, which is just the overtwist
$\wo\sigma$. Also the constant part of $\alpha$ is the extension
$\epsilon$. Thus neglecting the quadratic couplings to $\tau$ gives
that only the constant modes $\bar t_1,\ \bar t_2,\ \sigma$, and
$\epsilon$  respond to $\tau$ and $\sigma$, and they
minimize the function
\ifx\answ\bigans
\eqn\ezm{\eqalign{{\wo^2\over2}\biggl[&
\cbend'{\bar t_1}^{\ 2}
+\cbend\Bigl(\bar t_2-{\ctb\over\cbend}\sigma
-{\cbs\over\cbend}\epsilon\Bigr)^2
+\Bigl(\ctwist-{\ctb^2\over\cbend}\Bigl)\sigma^2
+\Bigl(\cstretch-{\cbs^2\over \cbend}\Bigr)\epsilon^2
+2\Bigl(\cts-{\ctb\cbs\over\cbend}\Bigr)\sigma
\epsilon\biggr]
\cr&
-\tau\epsilon - \Lambda\wo\sigma
\ .\cr}}
\else
\eqn\ezm{\eqalign{{\wo^2\over2}\biggl[&
\cbend'{\bar t_1}^{\ 2}
+\cbend\Bigl(\bar t_2\hbox{-}{\ctb\over\cbend}\sigma
\hbox{-}{\cbs\over\cbend}\epsilon\Bigr)^2
+\Bigl(\ctwist\hbox{-}{\ctb^2\over\cbend}\Bigl)\sigma^2
+\Bigl(\cstretch\hbox{-}{\cbs^2\over \cbend}\Bigr)\epsilon^2
+2\Bigl(\cts\hbox{-}{\ctb\cbs\over\cbend}\Bigr)\sigma
\epsilon\biggr]
\cr&
-\tau\epsilon - \Lambda\wo\sigma
\ .\cr}}
\fi
We see that $\bar t_2$, which is neither controlled nor observed,
adjusts slightly but that the measured twist-stretch coupling
$\bar\cts$ reflects the combination $\cts-{\ctb\cbs\over\cbend}$ of
intrinsic elastic parameters. Parenthetically we note that $\bar\ctwist
= \ctwist-{\ctb^2\over\cbend}$ is nearly equal to $\ctwist$ because we
expect $\ctb$ to be small, and similarly for $\bar\cstretch$. The
corrections are
small because they reflect the small deviation of DNA from a straight
circular rod.

We are almost done. Normally the linear force terms suffice since the
applied force $\tau/\wo^2$ is small compared to the persistence
length $\cbend$. The only exception to this rule comes from those
modes which
decouple completely from \ebtt\ when $\tau=0$: these modes will have
large fluctuations, and the nonlinear couplings to $\tau$ are needed
to cut them off. These dangerous modes are the Fourier modes of $t_1,\
t_2$ of wavenumber close to $\wo$, as we see from \ebtt. Their thermal
fluctuations indeed make a large contribution to $\exv{z\tot}$ at
small force. Fortunately, though, this contribution is completely
independent of $\sigma$, since $\sigma$ enters linearly in the problem
and hence makes no contribution to the fluctuation problem. Thus we can
simply absorb this contribution to $\exv{z\tot}$ into the constant
term of \eexpt.

Thus we have found the interpretation of the experimentally-determined
twist-stretch coupling found in section 2: in terms of the
intrinsic elasticity of DNA the slope in \eexpt\ fixes the combination
$(DA -GK)/AB$ to be 22~nm. The low-force stretching experiments give
bend stiffness\foot{This value for $\cbend$ is
slightly smaller than the traditional one. The authors of
ref.~\rBlock\
eliminated the electrostatic contribution to $\cbend$ by extrapolating to
high concentration of high-valence added salt.} $\cbend=40\,$nm
\rBlock. Other experiments fix $B,\ C$ to the approximate values
quoted earlier. The remaining two combinations of the six couplings in
\ebtt\ do not appear to be relevant for existing experiments.

\newsec{Microscopic Model}
The elastic theory in the previous section was very general, but it
gave no indication of the expected magnitudes of the various
couplings.
To gain further confidence in our result, we will now see how to
{\it estimate} the expected twist-stretch coupling based on the measured
values of the other elastic constants and geometrical information
about DNA. We will use a simple, intuitive microscopic picture of DNA
as a helical rod to show how twist-stretch coupling
can arise and get its general scaling with the geometric
parameters. While the model is unrealistic it captures the underlying
symmetries and shows that the value of $\bar\cts$ calculated above is
reasonable.

Our picture will be a  beam of isotropic elastic material of circular
cross-section, initially bent into a helix of pitch $\ell_0$,
with the beam center slightly displaced from the helix axis by
$d_0\ll\ell_0$ \rKrenk. \tfig\fone b defines notation (for realistic
depictions of DNA structure see for example \rCalla). In this section
we will  consider only uniform deformations of the helix; in
particular the helix axis will always be straight. It proves
convenient to define slightly different variables from the previous
section: instead of following the helix axis, now our curve follows
the centerline of the beam.  We again call the tangent to this curve
$\TTZ(s)$, where the
arc length $s$ runs from 0 to $\tilde L$, but now $\tilde L$ is
slightly longer than the end-to-end length of the relaxed beam.
Next we draw a second curve, the locus of points farthest from
the helix axis. Let
$\EEZ(s)$ be the field of vectors perpendicular to the tangent $\TTZ(s)$
and pointing from the first to the second curve. Finally complete
$\TTZ,\ \EEZ$ to an orthonormal triad by defining
$\FFZ=\TTZ\times\EEZ$.

The distorted beam will then have its modified
frame $\{\EE(s),\ \FF(s),\ \TT(s)\}$, where now $s$ is the arc length
rescaled by $(1+\alpha)\inv$ to again run from 0 to $\tilde L$ and
$\alpha$ is the axial strain as before. We also define strain
variables $\Omega_i$ as before; for the uniform
deformations considered these are constants independent of $s$.
For a nearly-straight, circular rod the elastic energy is then~\rLandau
\eqn\etsbm{f_3=\half\left[
\cbend(\Omega_2-\Omega_{20})^2+
\ctwist(\Omega_3-\Omega_{30})^2+
\cstretch\wo^2\alpha^2
\right]\ .}
We have rotated our reference frames about the tangent vector to
eliminate $\Omega_1$. The constants $\cbend,\ \ctwist$, and
$\cstretch$ can in
turn be expressed in terms of the effective Young modulus, shear modulus, and
diameter of the rod, but we instead use the measured values quoted earlier.
%the intrinsic bend stiffness is\ and $\ctwist=75\,$nm, $\cstretch=78\,$nm.

To use \etsbm\ we need to find $\Omega_2$ and $\Omega_3$ in terms of
the helix parameters:  helix axis offset $d$,
end-to-end length $z\tot$, and total rotation of the
cross-section. The latter quantity plays the role of linking number
for open DNA, and so we will call it Lk.
To get the required relations it is helpful
to use the physical image of a gyroscope rotating at ``angular frequency''
$|\vec\Omega|$ about an axis parallel to $\vec\Omega$ while moving at
constant ``speed'' along an axis $\TT$ fixed in the body.
%Thus in this analogy arc length plays the role of time.
We take ``time'' to run from 0 to $\tilde L$, the original relaxed arc
length; to allow for intrinsic stretching we then take the ``speed''
to be $1+\alpha$.
One then finds that
\eqn\ehelix{d=\Omega_2(1+\alpha)/|\vec\Omega|^2\ ,\quad
z\tot=\tilde L\Omega_3(1+\alpha)/|\vec\Omega|\ ,\quad
{\rm Lk}=\tilde L|\vec\Omega|/2\pi\ .}
To fix $\vec\Omega_0$ we impose the values $\alpha_0=0$, $\wo
=2\pi\lk_0/z_{\rm tot,0}=1.85/$nm, and a helix offset $d_0$. We will
choose $d_0$ to get the observed value of $\bar\cts$ and see that it is
reasonable. Working to second order in $d_0$ \ehelix\ gives
$\Omega_{30}=\wo(1-d_0^2\wo^2)$,
$\Omega_{20}=\do\wo^2$.

We must now minimize \etsbm\ with the constraint of fixed $z\tot$ and
\lk. Let $z\tot\equiv(1+\epsilon)z_{{\rm tot},0}$, so that
$\epsilon$ again measures changes in end-to-end distance, and
$\lk=(1+\sigma)\lk_0$ as usual. Again using \ehelix, one finds
\eqn\emini{\Omega_2-\Omega_{20}=\beta\wo\ ,\quad
\Omega_3-\Omega_{30}=\wo\sigma-\wo^2\do\beta\ ,\quad
\alpha=\epsilon+\wo\do\beta-(\wo{d_0})^2\sigma\ ,}
where $\beta$ is free. Substituting into \etsbm\ and minimizing over
$\beta$ reveals a $\sigma\epsilon$ coupling corresponding to
$\bar\cts= (\wo\do)^2(\ctwist-\cbend)\cstretch/\cbend$. This formula
fits our measured value of $\bar\cts$ if we choose $\do=0.2\,$nm.

The value of $\do$ is not known {\it a priori,} since of course DNA is
not really an elastic continuum with circular cross-section.
Nevertheless, inspection of the known molecular structure indeed
suggests an elastic center offset from the helix axis by a couple of
\AA ngstroms \rCalla. In any case we have shown that the measured value
of $\bar\cts$ is of the order of magnitude expected from simple elasticity
theory.\foot{Actually a helical beam can have a twist-stretch coupling
even if its axis is on center, $d_0=0$, provided its cross-section is
not circular~\rKrenk. To explain the observed coupling in this way
would require a rather large eccentricity of 70\%. However for
molecules such as actin, for which $d_0=0$, this second mechanism may
be important.}

\newsec{Conclusion}
We have pointed out a strong twist-stretch coupling in
torsionally-constrained DNA stretching experiments, evaluated it,
argued that it reflects intrinsic elasticity of the DNA duplex, and
shown that the value we obtained is consistent with elementary
considerations from classical elasticity theory. A greater challenge
remains to predict this coupling from the wealth of available
crystallographic information on the conformation of short oligomers.

{\frenchspacing
\ifx\prlmode\testp\else\vskip1truein \leftline{\bf Acknowledgments}
\noindent\fi
We would like to thank
D. Bensimon, S. Block, and J. Marko for their help and for
communicating their
results to us prior to publication, and W. Olson for discussions. }
RK, TL, and CO were  supported in part by NSF grant DMR96--32598.
PN was supported in part by NSF grant
DMR95--07366.
\listrefs
%\ifx\figflag\figI\else\listfigs\fi
\bye